\documentclass[12pt]{iopart}
\usepackage{graphicx}
\usepackage{subfigure}
\usepackage{epstopdf}
\usepackage{amssymb}
\usepackage{amsbsy}
\usepackage{color}
\begin{document}

\title[Collective Phenomena in the  LiHo$_x$Y$_{1-x}$F$_4$
Quantum Ising Ferromagnet]
{Collective Phenomena in the  LiHo$_x$Y$_{1-x}$F$_4$ 
Quantum Ising Magnet: Recent Progress and Open Questions}
\author{Michel J.P. Gingras$^{1,2}$, Patrik Henelius$^3$}
\address{$^1$Department of Physics \& Astronomy, University of Waterloo, Waterloo, Ontario, N2L 3G1, Canada} 
\address{$^2$Canadian Institute for Advanced Research, 180 Dundas Street West, Suite 1400, Toronto, Ontario, M5G 1Z8, Canada}
\address{$^3$Department of Theoretical Physics, Royal Institute of Technology, SE-106 91 Stockholm, Sweden}

\begin{abstract}
  In LiHo$_x$Y$_{1-x}$F$_4$, the magnetic Holmium Ho$^{3+}$ ions behave as
  effective Ising spins that can point parallel or antiparallel to the
  crystalline $c$-axis.
The predominant inter-Holmium interaction is dipolar, while
  the  Y$^{3+}$ ions are
  non-magnetic.  The application of a magnetic field $B_x$ transverse
  to the $c$-axis Ising direction leads to quantum spin-flip
  fluctuations, making this material a rare physical realization of
  the celebrated {\it transverse field Ising model}.  The problems of
  classical and transverse-field-induced quantum phase transitions in
  LiHo$_x$Y$_{1-x}$F$_4$ in the dipolar ferromagnetic ($x=1$), diluted
  ferromagnetic ($0.25 \lesssim x < 1$) and highly diluted $x\lesssim
  0.25$ dipolar spin glass regimes have attracted much experimental and
  theoretical interest over the past twenty-five years.  Two questions have
  received particular attention: (i) is there an {\it antiglass}
  (quantum disordered) phase at low Ho$^{3+}$ concentration and (ii)
  what is the mechanism responsible for the fast $B_x$-induced
  destruction of the ferromagnetic ($0.25 \lesssim x < 1)$ and spin
  glass ($x \lesssim 0.25$) phases?  This paper reviews some of the
  recent theoretical and experimental progress in our understanding of
  the collective phenomena at play in LiHo$_x$Y$_{1-x}$F$_4$, in both
  zero and nonzero $B_x$.
\end{abstract}

\date{\today}


\maketitle

\section{The pure  LiHoF$_4$ material}

\subsection{Dipolar ferromagnetism}

\label{Sect:dipfm}

\begin{figure}
\begin{center}
\subfigure{
\includegraphics[scale =0.6]{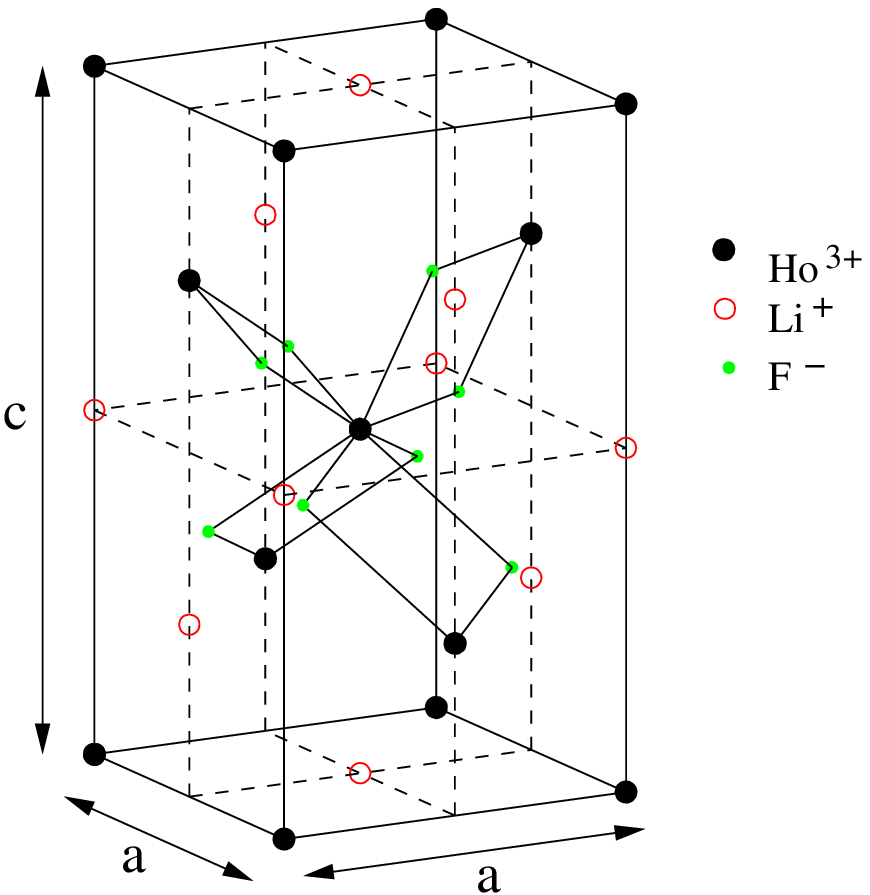}
\label{fig1a}
}
\subfigure{
\includegraphics[scale=0.5]{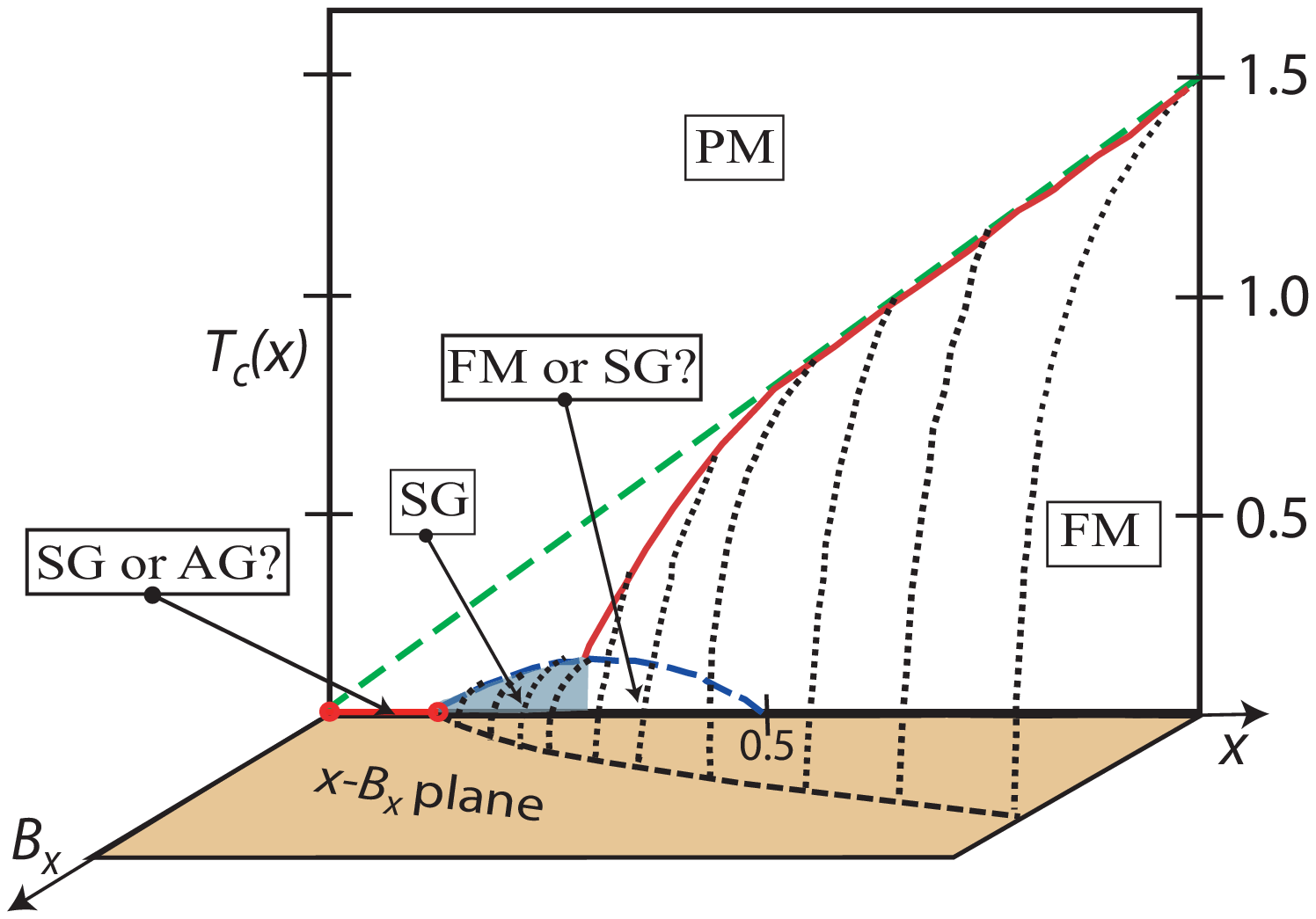}
\label{fig1b}
}
\label{fig1}
\end{center}
\noindent 
\caption{The left panel shows the crystalline tetragonal structure
of LiHoF$_4$ with lattice spacing 
$a=5.175$ $\mathring{A}$ and $c=10.75$ $\mathring{A}$.
The thin solid lines illustrate the superexchange pathways
between magnetic Ho$^{3+}$ ions (black circles) 
mediated via fluorine F$^-$ ions (green circles).
In LiHo$_x$Y$_{1-x}$F$_4$, the magnetic 
Ho$^{3+}$ ions are substituted randomly
by non-magnetic Y$^{3+}$ ions, with the lattice
structure remaining the same for all $x$.
The global Ising direction is along the $c$ axis
and a magnetic field $B_x$ is applied perpendicular
to that axis in transverse field experiments.
The right panel shows a schematic 
temperature ($T$) $-$ dilution ($x$) $-$ transverse
field ($B_x$) phase diagram of 
LiHo$_x$Y$_{1-x}$F$_4$.
At high temperature $T$ or large $B_x$, the system
is in a paramagnetic (PM) phase.
For $x=1$, LiHoF$_4$ is an Ising dipolar ferromagnet (FM) below
$T_c=1.53$ K. 
Upon the substitution of Ho$^{3+}$ by Y$^{3+}$, the FM 
phase persists down to $x \sim 0.25$. 
Upon cooling the diluted FM for $0.25\lesssim x \lesssim 0.5$, magnetic
susceptibility measurements find the development of a ``ferroglass''
regime (FM or SG?).
As $x$ decreases, random frustration builds in and,
for $x \lesssim 0.25$, a dipolar Ising spin glass (SG) phase develops
(shaded blue region).
Whether the dipolar SG exists down to $x=0^+$ or  an exotic
quantum disordered ``antiglass'' (AG) phase occurs at $x>0$ (red line segment)
is not yet resolved and is the subject of much controversy.
The application of a transverse field $B_x$ introduces quantum fluctuations
causing a reduction of $T_c(x)$ (surface delineated by dotted lines) which,
ultimately,  drives 
a zero temperature quantum phase transition at zero temperature
(dash-line curve in the $x-B_x$ plane).}
\end{figure}

The insulating rare-earth compound LiHoF$_4$ has the scheelite
structure depicted in Fig. 1 \cite{wyck65}.  In this material, only
the Holmium ions, Ho$^{3+}$, are magnetic.  LiHoF$_4$ forms a
tetragonal structure (space group C$_{4\rm h}^6-I4_1/a$) with lattice
constants $a=5.175$ $\mathring{A}$ and $c=10.75$ $\mathring{A}$.
There are 4 magnetic Ho$^{3+}$ ions per unit cell with fractional coordinates
$(0,0,\frac{1}{2})$, $(0,\frac{1}{2},\frac{3}{4})$, 
$(\frac{1}{2},\frac{1}{2},0)$ and $(\frac{1}{2},0,\frac{1}{4})$. 
In Fig. 1, the lines connecting the central Ho$^{3+}$ ion with neighbouring F$^-$ ions
and then to the nearest  Ho$^{3+}$ ions illustrate the superexchange pathways between
nearest-neighbour Ho$^{3+}$ cations \cite{menn84}.  As in most
magnetic rare-earth compounds, the unfilled $4f$ electronic orbitals
of the Ho$^{3+}$ ions are not very spatially extended. This makes
the exchange interaction weak and causes magnetostatic dipolar
couplings to be the strongest inter-ionic interactions.  Indeed,
LiHoF$_4$ is a dipolar Ising ferromagnet 
with a magnetic moment of  $\sim 7\mu_{\rm B}$ per Ho$^{3+}$ ion \cite{menn84}
 and with a critical temperature
$T_c \approx 1.53$ K \cite{menn84}.  In the mid-seventies,
crystal-field calculations \cite{hans75}, susceptibility \cite{cook75}
and spectroscopic measurements \cite{batt75} showed that LiHoF$_4$ is
an Ising-like dipolar ferromagnet with a highly anisotropic $g$-tensor
($g_{\perp} \approx 0$ and $g_{\parallel} \approx 14$). The dipolar
interaction has a long and interesting history and this physical
realization rekindled interest in dipolar ferromagnetism. 
Luttinger and Tisza had found that the ground state of a dipolar ferromagnet
depends on both the lattice structure and
sample shape \cite{lutt46}. However, taking the demagnetization field
into account, Griffiths showed that, in the absence of an
applied magnetic field, the free energy is independent of the sample
shape \cite{grif68}. Using renormalization group arguments, Larkin and
Khmelnitskii \cite{lark69} and Aharony \cite{ahar73} showed that
$d^*=3$ is the upper critical dimension for mean-field critical
behaviour in a dipolar ferromagnet.  
This observation led to significant experimental efforts to find the expected logarithmic
corrections to mean-field theory in LiHoF$_4$ \cite{grif80,beau78,ahle75,frow79,NikEll01}.  
The demagnetization field also renders the ferromagnetic transition quite peculiar
in the presence of
dipolar interactions \cite{cook75} as these oppose a uniform magnetization and leads 
to the formation of domains below the transition \cite{batt75,meye89}. 
The size and shape of the domains depend on an energy balance between surface and bulk
contributions \cite{kitt49}. For LiHoF$_4$, the ground state domain
configuration is expected to consist of parallel antiferromagnetically aligned 
sheets \cite{gaba84,gaba85,bilt09}. The size of the domains increases
with decreasing demagnetization factor and, in the limit of infinitely
long thin samples, a state of uniform magnetization is possible. 
 Also, for thick samples, a branched pattern  is predicted in the
domain structure  for a dipolar Ising ferromagnet \cite{gaba84,gaba85}.
LiHoF$_4$ is an excellent test case for theories of domain formation
due to the availability of an accurate microscopic model of the material (see below).
It remains an open experimental challenge to verify the nature of the
predicted rich domain structure.

\subsection{Microscopic model and transverse field Ising model for LiHoF$_4$}

The magnetic properties of LiHoF$_4$ are determined by the 4$f^{10}$
electrons of the Ho$^{3+}$ ions. Applying Hund's rules results in a 
17-fold degenerate $^5I_8$ electronic ground state. The crystal field
from the surrounding ions, described by a crystal field potential
$V_{\rm CF}$, lifts this degeneracy and gives rise to a two-fold
non-Kramers degenerate ground state, $\vert \psi_0^\pm \rangle$, and
an excited singlet, $\vert \psi_{\rm e}\rangle$, at an energy of
approximately 10 K above $\vert \psi_0^\pm \rangle$. 
The minimal microscopic Hamiltonian in the presence 
of a transverse magnetic field $\vec B=B_x\hat x$ 
can be written as \cite{chak04}:
\begin{eqnarray}
  H&=&\sum_i V_{\rm CF} (\vec{J}_i) - g_L\mu_B\sum_iB_{x}J_i^x
  +\frac{1}{2}(g_L\mu_B)^2\sum_{i\ne
    j}\mathcal{L}_{ij}^{\mu\nu}J^{\mu}_iJ^{\nu}_j \nonumber\\
  &+&{\rm J}_{\rm{ex}}\sum_{\langle i,j\rangle }
\vec{J_i} \cdot \vec{J}_j + A\sum_i \vec{I}_i \cdot \vec{J}_i ,
\label{firstprin}
\end{eqnarray}
where $\mu,\nu=x,y,z$.  $\mathcal{L}_{ij}^{\mu\nu}$ is the magnetic
dipole interaction,
$\mathcal{L}_{ij}^{\mu\nu}=
[\delta^{\mu\nu}|\vec{r}_{ij}|^2 - 3(\vec{r}_{ij})^{\mu}(\vec{r}_{ij})^{\nu}]/|\vec{r}_{ij}|^5$,
J$_{\rm{ex}}$ is the nearest-neighbour exchange interaction.   
$A$ is the strength of the hyperfine
interaction and $\vec{I}_{i}$ ($I=7/2$) is the total angular
angular momentum vector of the Ho-nucleus at the $i$-th site.
Disregarding the hyperfine interactions, an effective spin 1/2 model for LiHoF$_4$ 
from the above microscopic model of Eq.~(\ref{firstprin}) was derived in
Ref.~\cite{chak04} and revisited in Ref.~\cite{tabe08}.  
Without hyperfine interactions, the single site Hamiltonian (first two
terms in Eq.~(\ref{firstprin})) can be diagonalized numerically for arbitrary value of
the transverse field $B_x$.  The ground state doublet is split by
$B_x$, with an energy $\Delta(B_x$) between the two states.
Meanwhile, the excited $\vert \psi_{\rm e}\rangle$ singlet state
remains more than 10 K above the split doublet.  
Consequently, the $\vec J_i$ operators can be
projected onto the two-dimensional subspace of the two lowest energy eigenstates
via the relationship
$J_i^{\mu}=C_{\mu 0} +\sum_{\nu=x,y,z}C_{\mu\nu}(B_{x})\sigma_i^{\nu}$, 
hence providing a description in terms of effective pseudospin-1/2 operators.
The strongest effective interaction is $\propto
{(C_{zz})}^2\mathcal{L}^{zz}_{ij}\sigma^{z}_{i}\sigma^{z}_{j}$ and, to
an accuracy of a few percent \cite{chak04,tabe08}, 
the effective model can be written (up to a field dependent constant) as
\begin{eqnarray}
  H_{\rm{Ising}}&=& -\frac{1}{2} \Delta \sum_{i}\sigma^{x}_{i} 
   +\frac{1}{2}(g_{L}\mu_{B}C_{zz})^{2}\sum_{i\ne j}
\mathcal{L}^{zz}_{ij}\sigma^{z}_{i}\sigma^{z}_{j}
+ J_{\rm ex}(C_{zz})^{2} \sum_{\langle {i,j}\rangle} \sigma^{z}_{i}\sigma^{z}_{j}.
\label{isingfirst}
\end{eqnarray}
Note that the $C_{\mu \nu}$ parameters and the {\it effective}
transverse field $\Gamma$, $\Gamma\equiv \Delta/2$, depend on the applied physical transverse
field $B_x$.  In the limit of small $B_x$, $\Delta(B_x) \propto B_x^2$
\cite{chak04,tabe08}. 
In Eq.~(\ref{isingfirst}), $\Delta/2$ plays the role of an effective transverse field
acting on the $x$-component of the pseudospin $\vec \sigma_i$ $-$ hence the
realization of a transverse field Ising model (TFIM) in LiHoF$_4$ \cite{bitk96}.

For $B_x=0$ ($\Delta=0$), the effective dipolar model (\ref{isingfirst}) has been
analyzed using mean-field theory \cite{menn84,cook75} and classical
Monte Carlo simulations. Numerical Monte Carlo simulations are
complicated by the difficult nature of the conditionally-convergent
dipolar lattice sums.  An early Monte Carlo simulation using free
boundary conditions for rather small system sizes found a critical
temperature of 1.89 K \cite{jens89}.  
To obtain better accuracy with long range dipolar interactions, 
 there are two standard approaches that can be implemented: Ewald summation
\cite{bilt09,tabe08,xu91,xu92,bilt07,bilt08} and cavity methods
\cite{chak04,kret79}.  An early 
Monte Carlo simulation using the Ewald summation
method to a dipolar Ising ferromagnet on a body-centered lattice was
carried out by Xu {\it et al.} \cite{xu91,xu92}. 
Monte Carlo simulations using the
cavity method \cite{chak04} and the Ewald method
\cite{bilt09,tabe08} have determined the critical temperature for the dipolar
model  in Eq.~(\ref{isingfirst}) with $\Delta=0$
on the LiHoF$_4$  lattice and estimated the nearest-neighbour
(antiferromagnetic) exchange interaction J$_{\rm ex}$ needed to lower the
$T_c$ Monte Carlo value
in order to match it to the experimental value of 1.53 K \cite{menn84}.  
A more recent study, using
the Ewald summation method and system sizes up to 32 000 dipoles, found
that the magnetization, specific heat and susceptibility do
match experimental results at a quantitative level \cite{bilt09}. 
The pure LiHoF$_4$ material can therefore be viewed as one of the best
realizations of a  dipolar Ising model.
A review of materials that behave like Ising systems can be found in
Ref.~\cite{wolf00}.

\subsection{Hyperfine interaction}

As in other Ho-based materials \cite{krus69,bram01}, there is a strong
 hyperfine coupling ($A=0.039$ K) between the electronic ($J=8$) and nuclear ($I=7/2$)
moments in LiHoF$_4$.  The hyperfine contribution dominates the
specific heat below 0.5 K \cite{menn84}, and leads to a substantial
increase in the critical transverse field at low temperatures \cite{bitk96}. 
As we shall discuss in Section \ref{Sec:TFRF}, the hyperfine interaction 
also strongly influences the
phase diagram of the dilute compound LiHo$_x$Y$_{1-x}$F$_4$ in a
transverse field \cite{sche05,sche08}.  The hyperfine interaction is therefore
important and may not be omitted in a proper quantitative
description of the low-temperature properties of
LiHo$_x$Y$_{1-x}$F$_4$.

\subsection{Transverse field effects}

At a continuous phase transition, the order parameter fluctuates
coherently over increasing length scales as the critical point is
approached. At finite temperatures, the fluctuations are classical in
nature sufficiently close to the critical temperature.  In some
systems, at temperatures near absolute zero, a phase transition can be
induced by tuning an external parameter such as a magnetic field, pressure,
doping or disorder.  In such a case, the fluctuations are quantum mechanical
in nature and the transition is called a {\it quantum phase
  transition} (QPT) \cite{sond97,sach99}.  The archetypical model for
a quantum phase transition is the {\it transverse field Ising model}
(TFIM) \cite{genn63, youn75}.  There are, however, not many real Ising
magnets with experimentally accessible QPTs \cite{Stasiak} since the
required transverse field usually far exceeds laboratory fields
(currently less than 50 T for DC fields).
This would be the case for
dipolar Ising spin ice materials \cite{Ruff05},
but not for the Dy(OH)$_3$ and Ho(OH)$_3$  dipolar
Ising ferromagnets \cite{Stasiak}.
Due (i) to the weak dipolar
interaction and (ii) the low-lying singlet $\vert \psi_{\rm e}\rangle$ above
the ground Ising doublet $\vert \psi_0^{\pm}\rangle$,
LiHoF$_4$ is one of very few magnetic systems where one
can observe the vanishing of the magnetization as a transverse field
is applied \cite{bitk96}.  Interestingly, recent neutron scattering
measurements of LiHoF$_4$ have revealed that the relatively strong
hyperfine interactions in Ho mask the true QPT of the hybridized
electronuclear critical mode, and the energy gap in the solely
electronic degrees of freedom remains finite as the system passes
through the QPT \cite{ronn05}.

Given the accuracy of the effective Ising model for LiHoF$_4$ in the absence
of a transverse field, the material would seem to constitute an ideal
testing ground for the inclusion of non-commuting quantum-mechanical
terms in realistic  spin models of magnetic materials. 
 As outlined above,
there is a well defined procedure to describe the effect of the transverse field $B_x$
in LiHoF$_4$  within a TFIM  \cite{chak04,tabe08}.
However,  a quantum Monte Carlo calculation \cite{chak04} 
of the critical temperature, $T_c$, vs $B_x$ phase diagram reveals only a qualitative
agreement with the experiment of Ref.~\cite{bitk96}.  
In the low temperature region (below 0.5 K),
the effect of the hyperfine interactions on the $B_x-T$ phase diagram
becomes important, and this is only 
approximately taken into account in the simulation of Ref.~\cite{chak04}.  
Recent theoretical work proposes an alternative way to incorporate the effect
of the hyperfine interactions in an effective model of LiHoF$_4$ in
nonzero $B_x$ \cite{sche08}. 
Notwithstanding the importance of hyperfine effects at low temperature,
one notes that {\it even} close to the
classical critical temperature of $T_c=1.53$ K, the experimental
\cite{bitk96} phase boundary rises much more steeply than in the model of
Eq.~(\ref{isingfirst}) \cite{chak04,tabe08}.
 To compound the puzzle, a very recent
experiment using dilatometry \cite{Dunn} confirms the
previous $T_c(B_x)$ experimental phase diagram obtained on the basis of
susceptibility measurements \cite{bitk96}.  The reason for this
stability of the ordered state with respect to weak $B_x$  is  an open question.
  Possible explanations include magnetostriction effects, 
small terms that are neglected in the effective model of Eq.~(\ref{isingfirst})
(see Refs.~\cite{chak04,tabe08}
for a discussion of those terms) or
anisotropic exchange terms, such as ${{\rm J}^{uv}_{\rm ex}} (r_{ij}) J_i^u J_j^v$,  
altogether ignored in the microscopic model of Eq.~(\ref{firstprin}). 
While the latter interactions may largely be inconsequential at
the classical transition, when projected within the manifold of
$\vert \psi_0^\pm \rangle$ doublets when $B_x=0$ \cite{chin08}, these could
perhaps end up affecting the $T_c(B_x\ne 0)$ phase diagram when $B_x\ne 0$.

\section{LiHo$_x$Y$_{1-x}$F$_4$ $-$ transverse field and random field physics}
\label{Sec:TFRF}

\subsection{Disordered magnetic systems}

All real materials contain a certain amount of frozen-in random disorder such
as impurities, interstitials and crystalline defects.  One may then
ask how the thermodynamic properties and the stability of the ground state
of an otherwise idealized pure system are affected by various types of random
disorder \cite{grinstein84,fisher88}.  Since the mid 1970s, magnetic
systems have proven to be almost ideal paradigms to study the role of
random disorder in condensed matter systems. Examples of cornerstone
questions in the field of random disordered systems are: Does a
transition that is first order in a pure system remains first order in
the presence of random disorder \cite{Imry-Wortis}?  For a second order
transition, does an arbitrary small amount of random disorder change
the critical exponents and hence modify the universality class
\cite{Harris-crit}?  Does the long range order exhibited by a pure
system survive arbitrary small {\it random fields} conjugate to the
order parameter \cite{Imry-Ma}?  When the disorder is large enough and
the interactions are in strong competition and lead to {\it random
  frustration}, is there a thermodynamic transition at nonzero
temperature to a {\it spin glass} state \cite{mydosh,ISG}?  Under what
conditions does a random magnet exhibit nonanalytical behavior above
its critical temperature, in the so-called Griffiths phase, which is
a remnant of the long range order phase  displayed by  the parent
disorder-free system \cite{Griffiths-phase}?  The above
questions were originally by and large investigated for classical
phase transitions at nonzero temperature. In the late 1980s, those
and similar questions  started to attract considerable interest 
in the context of quantum phase transitions \cite{sach99}. 
Here again, the LiHoF$_4$ material, or more precisely, its Y$^{3+}$
diamagnetically diluted LiHo$_x$Y$_{1-x}$F$_4$ variant, has proven
to be a highly interesting material to investigate quantum phase transitions in
a disordered systemt and tuned by an external parameter. 
As in the pure LiHoF$_4$ material, the
parameter tuning the level of quantum (spin flip) fluctuations in
LiHo$_x$Y$_{1-x}$F$_4$ is a magnetic field $B_x$ applied perpendicularly
to the Ho$^{3+}$ Ising spin direction.

\subsection{Transverse field and random field effects in
  LiHo$_x$Y$_{1-x}$F$_4$}

The LiHo$_x$Y$_{1-x}$F$_4$ compound forms a solid solution from the
pure dipolar ferromagnetic LiHoF$_4$ to the non-magnetic LiYF$_4$
without a change of the crystalline structure and, thanks to the close
ionic radius of Ho$^{3+}$ and Y$^{3+}$, LiHo$_x$Y$_{1-x}$F$_4$
displays a minimal dependence of the $a$ and $c$ lattice parameters on $x$.
The critical ferromagnetic transition temperature, $T_c(x)$, 
 decreases as the magnetic Ho$^{3+}$ ions are substituted
by non-magnetic Y$^{3+}$.  For $0.25 \lesssim x < 1$,
LiHo$_x$Y$_{1-x}$F$_4$ can be described as a {\it diluted
  ferromagnet} (see Fig. 1). As discussed in Section \ref{Sect:dipfm}, 
the main interactions between the Ho$^{3+}$ ions are magnetostatic
dipole-dipole interactions.  The dipolar interaction is inherently
frustrated since its sign, that of $\mathcal{L}_{ij}^{\mu\nu}$ in
Eq.~(\ref{firstprin}), depends on the orientation of the
magnetic moments with respect to
$\vec r_{ij}\equiv \vec r_j -\vec r_i$, the
position vector of ion $j$ relative to ion $i$.  Hence, as a
result of the random substitution Ho$^{3+}$ $\rightarrow$ Y$^{3+}$,
{\it random frustration} builds up as $x$ decreases until, for
$x\approx 0.25$, the diluted ferromagnetic state gives way to a {\it
  dipolar spin glass} phase (see. Fig. 1).  We note that there has
been much debate as to whether such a spin glass phase occurs for
$x\lesssim 0.25$ in LiHo$_x$Y$_{1-x}$F$_4$ \cite{ancona,jonsson}.
We proceed in this section  with the assumption that a spin glass phase 
does exist, at least for $0.15 \lesssim x \lesssim  0.25$.  
We return in Section \ref{Sec:SGAG} to the question of the existence 
of a thermodynamic spin glass phase in
LiHo$_x$Y$_{1-x}$F$_4$ for $x \lesssim 0.25$, below the concentration 
where the diluted ferromagnetic  phase has disappeared (see Fig. 1).

The essence of the research on transverse field induced quantum
fluctuations and quantum phase transitions in LiHo$_x$Y$_{1-x}$F$_4$
originally lay in the expectation that this material may be a
physical realization of a transverse field Ising model (TFIM) with
Hamiltonian
\begin{eqnarray}
H_{\rm TFIM} = -\sum_{i>j} \Omega_{ij}^{zz} \sigma_i^z \sigma_j^z 
-\Gamma \sum_i \sigma_i^x,
\label{tfim_rnd}
\end{eqnarray}
with effective transverse field $\Gamma$.  In Eq.~(\ref{tfim_rnd}),
the randomly frustrated $\Omega_{ij}^{zz}$ couplings model, or rather mimic,
the site-diluted 
$\mathcal{L}_{ij}^{zz}$ 
$\rightarrow$
$\epsilon_i \epsilon_j \mathcal{L}_{ij}^{zz}$ in Eq.~(\ref{isingfirst}), 
where $\epsilon_i=1$ when the site $i$ is
occupied by a magnetic Ho$^{3+}$ ion and $\epsilon_i=0$ when it is 
occupied by non-magnetic Y$^{3+}$. 
In theoretical studies, important simplifications have been made by
taking a Gaussian distribution for $\Omega_{ij}^{zz}$ of mean $\Omega_0$ and standard
deviation $\Omega$ and setting $\epsilon_i =1$
$\forall i$ \cite{Miller,Guo94,Guo96,KimKim}.  
Of particular interest in these studies
was the prediction of Griffiths phase physics near the QPT both in two
and three dimensional spin glass versions ($\Omega_0=0$)  of 
Eq.~(\ref{tfim_rnd}) \cite{Guo96}.

Two puzzles were identified when the behaviour of
LiHo$_x$Y$_{1-x}$F$_4$ subject to $B_x\ne 0$ was first compared with
expectations from theoretical and numerical studies.  Firstly, while $dT_c/dx
\propto x$ in the diluted ferromagnetic (FM) regime ($0.25 \lesssim x
< 1.0$, see Fig. 1), as in mean-field theory \cite{Brooke}, the rate
at which $T_c(B_{x})$ is reduced by $B_{x}$ gets progressively faster
than mean-field theory predicts as the Ho$^{3+}$ concentration $x$
decreases~\cite{Brooke-thesis}.  This means that, in comparison with
the energy scale for FM order that determines the classical 
transition at ${T_c(B_{x}=0,x)}$,
$B_x$ becomes more efficient at destroying FM order the lower the
concentration $x$ is \cite{Brooke-thesis}.  Secondly, for ${B_{x}=0}$,
LiHo$_{0.167}$Y$_{0.833}$F$_4$ has generally been believed to exhibit
a conventional spin glass (SG) transition \cite{ancona} (see however
Ref.~\cite{jonsson,jonsson_cmp}), with the transition signaled by a nonlinear
magnetic susceptibility, $\chi_3$, diverging at $T_g$ as
$\chi_3(T)\propto (T-T_g)^{-\gamma}$~\cite{mydosh}.  However,
$\chi_3(T)$ in LiHo$_{0.167}$Y$_{0.833}$F$_4$ becomes less singular as
$B_{x}$ is increased from ${B_{x}=0}$.
This suggests that, in  contrast with numerical work \cite{Guo94,Guo96}, 
no $B_x$-driven quantum phase transition between a polarized 
quantum paramagnetic state and a SG state occurs
as $T\rightarrow 0$~\cite{Wu,Wu-thesis}.

The resolution of these two puzzles lies in the observation that, in
the properly derived effective low-energy theory akin to
Eq.~(\ref{isingfirst}) when the occupation factor $\epsilon_i \ne 1$
 \cite{tabe06,tabe08rf}, the transverse field $B_x$ induces
not only quantum fluctuations, but also {\it random fields}
\cite{grinstein84,fisher88}, $h_i^z$, that couple linearly to the
$\hat z$ component of $\vec \sigma_i$
\cite{sche05,sche08,tabe06,tabe08rf,sche06,sche08rf}. 
 There are two key physical ingredients responsible for these random fields.
Firstly, the magnetic field $B_x$, perpendicular to the $c$-axis Ising
direction, induces a finite expectation value of the Ho$^{3+}$ magnetic
moments along $\hat x$, $\langle \mu_i^x \rangle = g\mu_B \langle
J_i^x \rangle$, at {\it all}  temperatures. Quantum mechanically, this
moment arises from the $B_x$-induced admixing of the (initially
strictly Ising) $\vert \psi_0^\pm\rangle$ ground doublet with the
excited crystal field states.  
Secondly, the bare microscopic Hamiltonian in Eq.~(\ref{firstprin}),
as opposed to the effective TFIM of Eq.~(\ref{isingfirst}) and its
simplified version in Eq.~(\ref{tfim_rnd}), contains off-diagonal
(anisotropic) $\mathcal{L}_{ij}^{\mu\nu}$ ($\mu\ne\nu$) terms of
dipolar origin that couple the various components $J_i^\mu$ of $\vec
J_i$.  In pure LiHoF$_4$, perfect mirror symmetries 
make the lattice sum $h_{i}^z\equiv \sum_{j}
\mathcal{L}_{ij}^{zx} \langle J_j^x \rangle$ vanish identically.  For diluted
LiHo$_x$Y$_{1-x}$F$_4$, $h_{i}^z$ no longer vanishes; the 
transverse magnetic moments
 $\langle  \mu _i^x \rangle$ induce, via the
random $\epsilon_i \mathcal{L}_{ij}^{zx} \langle J_i^x\rangle$ terms, 
internal random fields  that act on the $z$ component of $\vec J_i$, $J_i^z$.  In other words,
diluted LiHo$_x$Y$_{1-x}$F$_4$ subject to a transverse field $B_x$
maps onto a TFIM 
with the addition of {\it correlated random fields}~\cite{tabe06,tabe08rf}.  
For $0.25 \lesssim x < 1$,
this makes LiHo$_x$Y$_{1-x}$F$_4$ a unique example of a {\it random
 field Ising ferromagnet model}.  
Indeed, the experimental procedure for
creating such a system is convoluted and typically considers a diluted
Ising antiferromagnet in a uniform longitudinal field $B_z$ \cite{fishman,Cardy}.
Analytical \cite{sche08,tabe06,tabe08rf,sche06,sche08rf} and numerical
\cite{tabe06,tabe08rf,sche06} studies have shown, rather convincingly,
that the presence of $B_x$-induced 
random fields $h_i^z$ can explain the two aforementioned puzzles observed
in LiHo$_x$Y$_{1-x}$F$_4$ in nonzero $B_x$, in both the diluted FM
regime and in the SG regime. In particular, the $B_x$-induced random
fields $h_i^z$ eliminate the divergence of the nonlinear
susceptibility $\chi_3$   in the spin glass
regime $x \lesssim 0.25$ \cite{sche05,tabe06,sche06}.

\subsection{Open questions regarding the random field physics in
  LiHo$_x$Y$_{1-x}$F$_4$}

Because of space constraint, we have focused above on the role
of the induced random fields on the 
phase transitions that occur in the diluted ferromagnetic and spin glass regimes.
As the critical (ferromagnetic or spin glass)  temperature
decreases, and drops below approximately 0.5 K, the role of the
hyperfine interaction in inhibiting quantum fluctuations
and strongly renormalizing the critical transverse field cannot
be ignored \cite{sche05,sche08}. 
While progress has been achieved as to how to proceed \cite{sche08},
 work does remain to be done to investigate via quantitatively
accurate calculations (e.g. quantum Monte Carlo simulations) 
the role of hyperfine interactions on the global $x-T-B_x$ phase
diagram of  LiHo$_x$Y$_{1-x}$F$_4$.

In the context of the physics of induced random fields, it is worthwhile
to note that theoretical work predicts that in the simplest
TFIM of Eq.~(\ref{tfim_rnd}), 
the addition of uncorrelated random fields should
destroy the zero temperature quantum criticality and
drive the renormalization group flow towards the fluctuationless
(classical) zero temperature random field fixed point \cite{Senthil}.
It would be interesting to investigate the implication of that result
for the {\it correlated} $h_i^z$ random fields discussed above and that
exist in LiHo$_x$Y$_{1-x}$F$_4$ in nonzero $B_x$.
Finally, we note that the presence of induced random fields has
been invoked to interpret the transverse field and temperature dependence 
(e.g. scaling behavior) of the magnetic susceptibility in  LiHo$_x$Y$_{1-x}$F$_4$ ($x=0.44$).
In particular, these random fields have been
 postulated to give rise to Griffiths' singularity effects
in the paramagnetic regime at temperatures just above $T_c$ \cite{silevitch}.
It would be useful to  explore this claim further. 
In fact, a systematic experimental investigation of the fascinating physics
of random field systems in the diluted ferromagnetic regime of  
LiHo$_x$Y$_{1-x}$F$_4$ is probably warranted.

\section{LiHo$_x$Y$_{1-x}$F$_4$ $-$ spin glass vs antiglass}
\label{Sec:SGAG}

\subsection{Spin glass vs antiglass physics in LiHo$_x$Y$_{1-x}$F$_4$}

The previous section discussed the properties of LiHo$_x$Y$_{1-x}$F$_4$ in nonzero
transverse field $B_x$. We now turn to the question of the existence of a spin glass
phase in LiHo$_x$Y$_{1-x}$F$_4$ for $x\lesssim 0.25$ and in zero $B_x$.

The $x-T$ phase diagram of 
LiHo$_x$Y$_{1-x}$F$_4$ in the limit of large dilution ($x \lesssim 0.25$) is
a subject of much controversy.  Theoretical studies of the dilute
Ising dipolar model predict that a spin-glass state may be favored
over ferromagnetic order at high dilution \cite{xu91,step81}.  Since
there is no percolation threshold for  long range dipolar
couplings, a thermodynamic spin glass state is expected to persist all
the way to $x=0^+$ \cite{step81}.  Furthermore, a study of the quantum corrections to
the Ising model arising from interaction-induced virtual crystal field
fluctuations finds that these quantum effects should be small in
LiHo$_x$Y$_{1-x}$F$_4$ \cite{chin08}. 
For example, this is in contrast to the
Tb$_2$Ti$_2$O$_7$ and Tb$_2$Sn$_2$O$_7$ pyrochlore Ising magnets
\cite{molavian,mcclarty} and, therefore, are unlikely to destroy the spin glass
phase in LiHo$_x$Y$_{1-x}$F$_4$  at finite Ho$^{3+}$ concentration.
Consequently, in light of the
accumulated evidence for a thermodynamic spin glass phase transition in
the three-dimensional Edwards-Anderson (EA) Ising spin glass model \cite{ISG}, 
one would also expect such a transition in LiHo$_x$Y$_{1-x}$F$_4$ for
$0 < x \ll 1$ since the EA model and dipolar systems should
be in the same universality class for a spin glass transition \cite{braymooreyoung}.

Measurements of the imaginary part of the magnetic
susceptibility, $\chi^"$, in LiHo$_{0.045}$Y$_{0.955}$F$_4$ by Reich
{\it et al.} ~\cite{reic87,reic90} was the first to challenge the above
theoretical picture of a spin glass phase down to $x=0^+$.  
In a typical glassy system, the frequency
distribution of $\chi^"$ broadens upon lowering the temperature \cite{mydosh}.  In
contrast, the authors of Ref.~\cite{reic87,reic90} found a
distribution that {\it narrows} as the temperature is reduced,
indicative of an overall speed-up of the dynamics and an 
``antiglass'' (quantum disordered) ground state. 
The measurement was later repeated
\cite{ghos02} to lower temperatures (50 mK), still showing an evident
narrowing of the spectrum.  
The manifestation of dissipationless dynamics of a broad inhomogeneous distribution
of quantum oscillators within the inferred antiglass regime
was further demonstrated through 
(i) hole burning in the AC susceptibility spectrum \cite{ghos02}
and, 
(ii) perhaps most interestingly, via the observation
of persistent oscillations for up to 30 seconds after an external AC
field is turned off \cite{ghos02}.
It has been argued that this antiglass behaviour
may originate from a quantum mechanical
entanglement of the magnetic dipole
moments and the formation of a sort 
of random-singlet state \cite{ghos03}.
However, as mentioned above, the quantum effect corrections to the
otherwise perfect Ising  nature of the single-ion
crystal field ground doublet is expected to be very small
in LiHo$_x$Y$_{1-x}$F$_4$ \cite{chin08}. Also, while hyperfine interactions do
 strongly renormalize the critical transverse field in transverse field
experiments  (see Section 2) \cite{sche05,sche08}, at the same time,
preliminary calculations suggest that
these interactions may 
not be efficient at quenching the spin glass phase at a
Ho$^{3+}$ concentration of order of 5\% \cite{chin08}.
It is therefore not clear 
how the entanglement mechanism proposed in Ref.\cite{ghos03}
on the basis of an ad-hoc model can 
really proceed efficiently when starting from the microscopic model
of Eq.~(\ref{firstprin}).

Apart from the question of the existence of an antiglass phase, there
have also been contradictory experimental reports on the spin glass
transition in the regime $0.1 \lesssim x\lesssim 0.2$, where it was 
generally believed to be well established. 
 Specifically, J\"onsson {\it et al.} found that
the non-linear susceptibility $\chi_3$ does not diverge for samples
 with $x=4.5$\% and, surprisingly, for $x=16.5$\% as well.
These results prompted  J\"onsson {\it et al.} to conclude  
that there is no thermodynamic spin glass phase at all in the compound \cite{jonsson}.
In response to this work, Ancona-Torres {\it et al.} re-examined the
non-linear susceptibility for $x=16.7$\% and $x=19.8$\% 
and concluded that $\chi_3$ does diverge \cite{ancona}.
A possible explanation put forward in Ref.~\cite{ancona} 
for the discrepancy between the works of Ref.~\cite{jonsson} and Ref.~\cite{ancona}
is that the sweep rates of the magnetic field in the work
of J\"onsson {\it et al.} \cite{jonsson} 
were too fast and the fields too strong to expose 
the true spin glass critical behaviour of the system.
To compound the debate, J\"onsson and co-workers \cite{jonsson_cmp}
continue to maintain
that evidence for a spin glass transition in LiHo$_x$Y$_{1-x}$F$_4$ for
$x=16.5$\%  is inconclusive, unlike the claims of 
Ancona-Torres {\it et al.} \cite{ancona}.

Several numerical studies have aimed at finding a spin-glass
transition in  effective dipolar Ising models mimicking the dilute compound
LiHo$_x$Y$_{1-x}$F$_4$.  
The original work of Xu {\it et al.} only identified the disappearance
of the long range ferromagnetic order for a diluted dipolar
body-centered cubic lattice and did not 
investigate the spin glass freezing \cite{xu91}.  
The traditional way
to identify the glass transition has been to calculate the Edwards-Anderson
overlap between two replicas, \#1 and \#2, 
$q=1/N \sum_i \sigma_{i,1}^z\sigma_{i,2}^z$, 
and to determine the temperature at which the corresponding Binder ratios, 
$g=1-[\langle q^4\rangle] /3[\langle q^2\rangle]^2$,
for different system sizes intersect. However, several
Monte Carlo studies \cite{bilt07,bilt08} have been unable to
detect a spin glass transition on the basis of such intersection, or other
finite-size scaling properties of $q$ \cite{snid05}.
 These ``first generation'' Monte Carlo simulations of diluted dipole-coupled
Ising spins did therefore suggest that the absence of a
thermodynamic spin glass transition in LiHo$_x$Y$_{1-x}$F$_4$ at small
$x$ could possibly have a {\it classical} origin.  However, as we discuss in the
following section, more recent experimental and numerical 
 studies reveal a different picture.

\subsection{Recent developments in spin glass physics in LiHo$_x$Y$_{1-x}$F$_4$}

\begin{figure}
\begin{center}
\subfigure{
\includegraphics[scale=0.33]{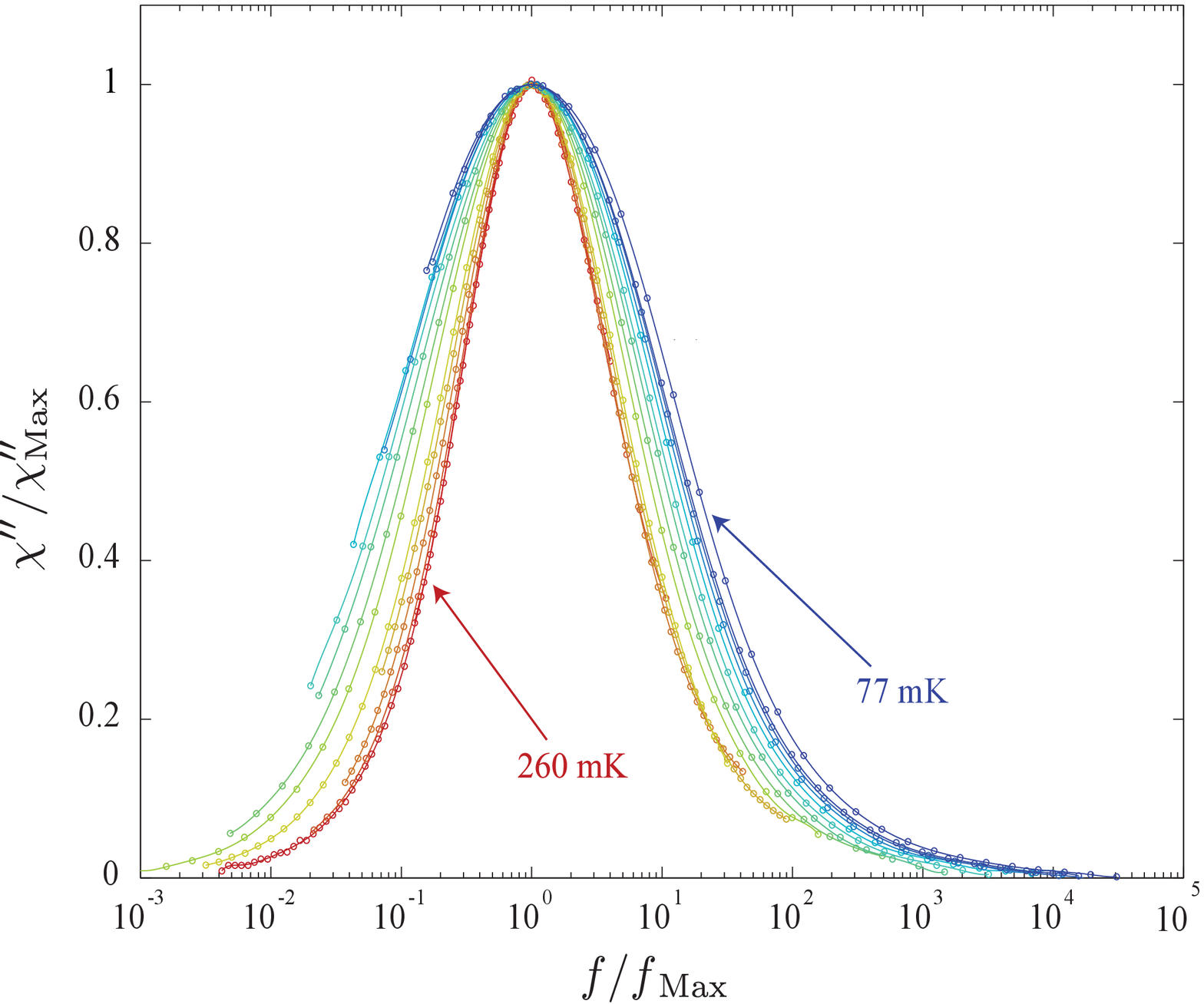}
\label{fig2a}
}
\subfigure{
\includegraphics[scale=0.33]{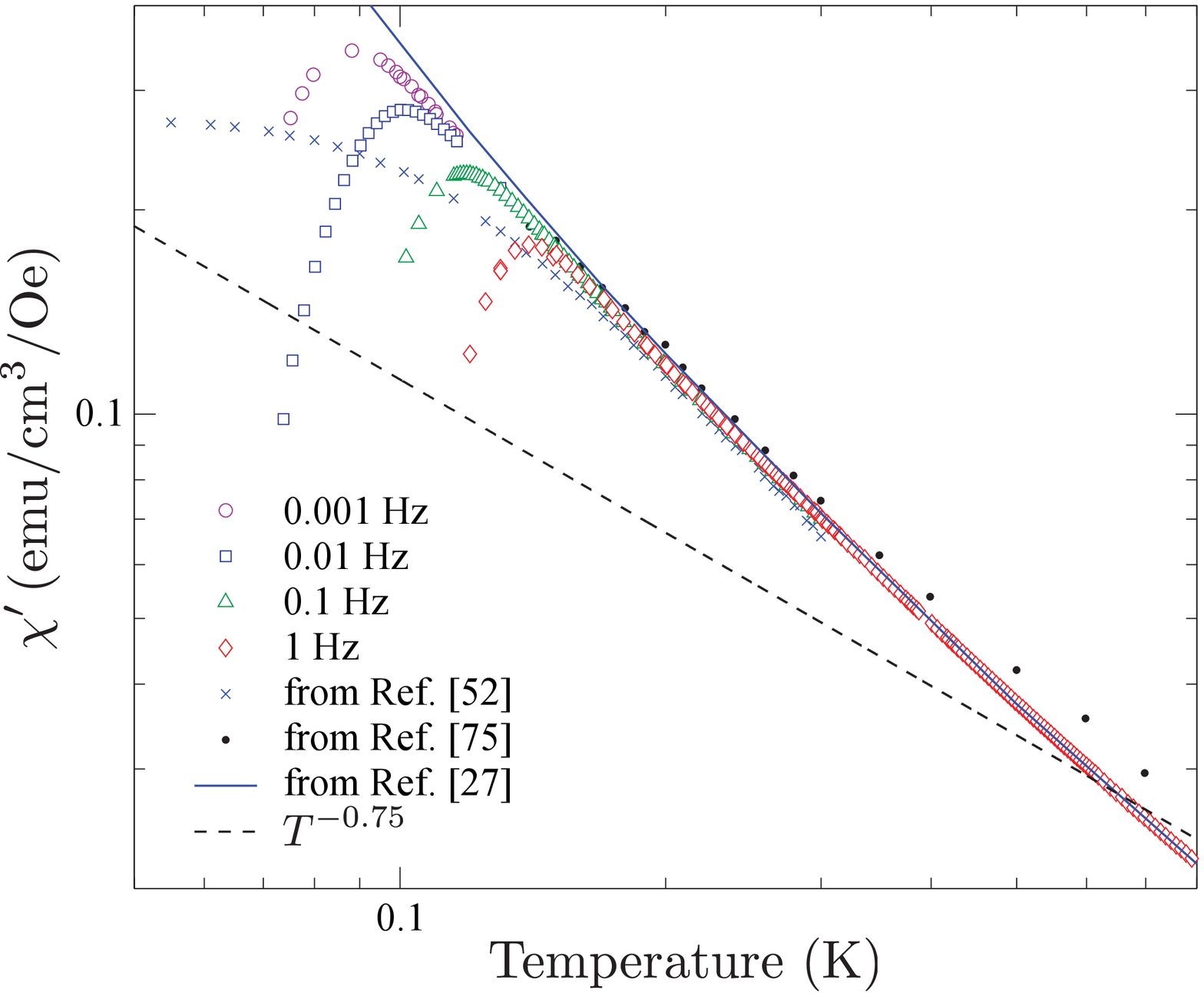}
\label{fig2b}
}
\label{fig2}
\end{center}
\noindent
\caption{The left panel shows a plot of the imaginary part of the
AC susceptibility, $\chi"$, as a function of frequency, $f$, 
normalized to its maximum value $\chi"_{\rm max}$ 
at frequency $f_{\rm max}$, and for temperatures between
77 mK and 260 mK.
The figure shows a broadening of the spectrum below and
above the peak frequency $f_{\rm max}$ at which $\chi"(f)$ peaks
(adapted from Ref.~\cite{quil08}).
The right panel shows the real part of the AC susceptibility,
 $\chi'(f)$, as a function of temperature. One notes the general
good agreement with the $\lim_{f\rightarrow 0} \chi"(f)$ data set
from Ref.~\cite{reic90}
down to a temperature $\sim 180$ mk  as well
as with the DC susceptibility data determined via classical
Monte Carlo \cite{bilt08}. 
The figure, adapted from Ref.~\cite{quil08},
 illustrates the disagreement with the
results of J{\"o}nsson  {\it et al.} in Ref.~\cite{jonsson}
as well as with the $T^{-0.75}$ behaviour reported in Ref.~\cite{ghos03} and
argued to support an entangled collective singlet ground state.}
\end{figure}


Since much of the evidence for the antiglass phase is based on the
scaling of the imaginary part of the AC susceptibility, the
reproducibility of these measurements is very important. 
Recent measurements by Quilliam {\it et al.} \cite{quil08} contradict
the earlier results  \cite{reic87,reic90,ghos02} and find that the
frequency distribution of $\chi"$, 
{\it broadens} as the temperature is reduced also in the high dilution limit
($x=4.5$\%) as shown in the left panel of Fig. 2.
 Furthermore, the quasi-static susceptibility, $\chi_0$, 
reported in Ref.~\cite{quil08} does not show the unusual $T^{-0.75}$ 
divergence found earlier \cite{ghos03}.  
We note, perplexingly, as noted in 
Refs.~\cite{bilt08,quil08}, that $\chi_0$ reported by Ghosh {\it et al.} 
in Ref.~\cite{ghos03} does not agree
with the previous results \cite{reic87,reic90} by the same group on,  
presumably, the very same sample. 
Meanwhile, the experimental $\chi_0$ data of 
Quilliam {\it et al.} ~\cite{quil08} 
agree reasonably well with the early results of Refs.~\cite{reic87,reic90} down
to a temperature of approximately 180 mK, 
as well as with the DC susceptibility found in 
a {\it classical} Monte Carlo simulation of Eq.~(\ref{isingfirst})
with $B_x=0$ \cite{bilt08}, as shown in the right panel of Fig. 2.  
The work of Ref.~\cite{quil08}  would
appear to give a strong indication that there is conventional spin glass
freezing  in the high dilution limit of LiHo$_x$Y$_{1-x}$F$_4$. 
 Furthermore, the specific heat data
reported in Ref.~\cite{quil07} do not display any
of the unusual features that were previously interpreted 
as a telltale signature of entangled magnetic moments 
within a quantum disordered antiglass state \cite{ghos03}.  
Note that the specific heat determined via
classical Monte Carlo simulations \cite{bilt08}
does not agree particularly well with either set of experimental data.
This discrepancy may be in part due to the difficulty in taking
into account accurately  of
the hyperfine contribution to the experimental specific heat data.

The reason for the difference between the AC susceptibility
and specific heat experimental results of
Refs.~\cite{reic87,reic90,ghos02,ghos03} and 
Refs.~\cite{quil08,quil07} remains an open question. 
Due to the very slow dynamics, all these measurements are
technically very demanding. However, we believe that it would be important to
settle the question of the nature of the high dilution phase in
LiHo$_x$Y$_{1-x}$F$_4$.
Perhaps the study of other highly diluted Ising magnetic materials could help
shed some light on the physics at play in LiHo$_x$Y$_{1-x}$F$_4$.
In this context, a study of the highly diluted Ising spin ice materials 
(Dy$_{x}$Y$_{1-x}$)$_2$Ti$_2$O$_7$ and
(Ho$_{x}$Y$_{1-x}$)$_2$Ti$_2$O$_7$ \cite{ke07} as well as the proposed
Ho$_x$Y$_{1-x}$(OH)$_3$ and Dy$_x$Y$_{1-x}$(OH)$_3$ TFIM materials \cite{Stasiak}
to compare with LiHo$_x$Y$_{1-x}$F$_4$ could prove interesting and instructive.

The failure of previous Monte Carlo simulations
\cite{bilt07,bilt08,snid05}
 to find a thermodynamic spin glass  transition in the dipolar
model would seem to have been explained recently.
As in three-dimensional Edward-Anderson Ising spin glass model \cite{ISG}, 
the spin glass correlation length $\xi_{\rm sg}$ turns out to be a much better indicator of
a spin glass  transition than the Binder ratio $g$.  A recent Monte Carlo study
\cite{tam09}  that focuses on  $\xi_{\rm sg}$  confirms that this
is also the case also for the dipolar Ising model.  In Ref.~\cite{tam09}, a finite-size
crossing of $\xi_{\rm sg}$ divided by the linear system size $L$, $\xi_{\rm sg}/L$,
is observed, providing good  evidence for a finite temperature
spin-glass transition, in agreement with the most recent AC
susceptibility experiments \cite{quil08}.
However, the critical temperature determined by the 
crossing of $\xi_{\rm sg}/L$ is slightly different for $\xi_{\rm sg}$ along the 
$a$ and $c$ axes.
This indicates that the finite size corrections to scaling are not under control for
the system sizes considered and  simulations on larger system sizes are therefore needed.
A very recent numerical study argues for a quasi-long-range ordered spin glass
phase in a system of diluted Ising dipoles ~\cite{alonso}.

\subsection{Open questions regarding the spin glass physics in LiHo$_x$Y$_{1-x}$F$_4$}

The single most important open question is the nature of 
the magnetic state in the limit of high dilution $(x < 0.1)$. 
Is it a traditional spin glass, as suggested by the recent measurements of 
Quilliam {\it et al.} \cite{quil08} or is it an unusual antiglass spin 
liquid \cite{reic87,reic90,ghos02,ghos03}? 
One may even ask whether there is a spin glass phase
 at all for $x\lesssim 0.25$ \cite{ancona,jonsson,jonsson_cmp}.
It would appear that the only way to resolve this issue is through more experiments. 
The ``sample quality'' (e.g. impurity, homogeneous Y$^{3+}$ disorder, etc) 
may be of  importance in this highly  dilute limit and this issue should be examined carefully.
Sample shape and  demagnetization effects should not affect the existence of the spin glass
phase, but this should nevertheless be investigated. 
There is a significant 
 qualitative difference between the static and dynamic susceptibility measurements 
reported by three research groups, and a new and independent series of
measurements could help shed light on this outstanding question.
It appears that the  nonlinear susceptibility results of Ref.~\cite{jonsson}
are being dismissed on
the basis of measurements that do not really probe the spin glass critical regime. 
However, disagreement on this interpretation remains \cite{jonsson_cmp}.
The hole-burning and persistent oscillations observed by  Ghosh {\it et al.}  \cite{ghos02}
are also indicative of a very unusual physical state, yet they have not been reported or
reproduced by another  group. 
To reiterate, we believe that at this stage,
in order for the field to move forward, more experiments that carefully
explore the critical regime, notwithstanding the highly technical burden associated with those,
are very much needed.

On the numerical side, it would be of interest to go beyond the classical Monte Carlo 
simulations that have so far been applied to the dilute system.
 The effects of the hyperfine interactions and off-diagonal interactions between
the $\vec \sigma_i$ pseudospins are challenging to
 incorporate in a quantum Monte Carlo simulation since they lead to a sign problem. 
Nevertheless, it would be important to invest some effort in this problem 
in order to achieve more quantitative comparisons with experiments.

\section{Conclusion}

Due to the existence of a well-established two-state Ising model 
for the rare-earth compound LiHo$_x$Y$_{1-x}$F$_4$,
this material constitutes a model 
magnet where  theory, experiments and simulations can be  compared
 and contrasted at a quantitative level. 
Despite the apparent simplicity of the underlying model 
 for the parent compound, the introduction of disorder through 
dilution and quantum effects through a transverse magnetic field 
leads to a complex system which continues to challenge experimentalists
 and theorists alike. The two foremost outstanding questions are 
whether a thermodynamic
spin glass phase survives in the limit of
high  dilution of the magnetic Ho$^{3+}$ ions 
and the reason for the unexpected stability 
of the ferromagnetic ground state of pure LiHoF$_4$ with respect 
to an applied transverse field. Continued research into this 
remarkable realization of the Ising model will hopefully continue 
to shed light on such diverse topics as random fields, 
coherent oscillations and tunable quantum fluctuations.


\vspace{1cm}

\bibliography{review}

\end{document}